\documentclass[a4paper,11pt]{article}
\pdfoutput=1 

\usepackage{jinstpub} 

\title{\boldmath X-Arapuca long term test}

\author[c]{V.~Andreossi,}
\author[d,a]{Z.~Balmforth,}
\author[c,b]{A.~A.~Bergamini~Machado,}
\author[c]{G.~Botogoske,}
\author[b]{N.~Canci,}
\author[c]{R.~de~Aguiar,}
\author[c]{P.~Duarte~De~Almeida,}
\author[a,b,1]{F.~Di~Capua,\note{Corresponding author}}
\author[a,b]{G.~Fiorillo,}
\author[b]{G.~Grauso,}
\author[a,b]{G.~Matteucci,}
\author[a]{S.~Ravinthiran,}
\author[c,b]{E.~Segreto,}
\author[a,b]{Y.~Suvorov}

\affiliation[a]{Physics Department, Universit\`a degli Studi di Napoli Federico II,\\Napoli 80126, Italy}
\affiliation[b]{INFN - Sezione di Napoli,\\Napoli 80126, Italy}
\affiliation[c]{Universidade Estadual de Campinas, UNICAMP, Brazil}
\affiliation[d]{Department of Physics, Royal Holloway University of London,\\Egham TW20 0EX, UK}

\emailAdd{dicapua@na.infn.it}

\abstract{The photon detection system of the DUNE experiment is based on the X-ARAPUCA light trap. The basic elements of the X-ARAPUCA are the dichroic filters coated with wavelength shifter (para-Therphenyl), a waveshifting plate and an array of SiPMs which detects the trapped photons. A small scale prototype of the X-ARAPUCA has been installed in liquid argon in a dedicated facility at INFN-Napoli and exposed to alpha particles from a source.
In order to test the stability of the overall device response the X-ARAPUCA was kept for 10 days in liquid argon continuously purified.
The performed tests allowed for a preliminary estimation of the X-ARAPUCA absolute photon detection efficiency.}


\proceeding{LIDINE2022\\
  September 21-23, 2022\\
  University of Warsaw Library}

\begin{document}
\maketitle
\flushbottom


\section{Introduction}
\label{sec:intro}

Next generation neutrino experiments will investigate new physics beyond the Standard Model,
addressing the measurement of the CP violating phase in the leptonic sector.
A significant contribution is expected to the completion of the understanding of the 
standard neutrino oscillation picture by measuring the mixing parameters and  the neutrino mass hierarchy. \\
The Deep Underground Neutrino Experiment (DUNE)~\cite{Dune} on the Fermilab Long-Baseline Neutrino Facility (LBNF) represents one of the most relevant experiments in this field.
LBNF
provides an high intensity, broad band neutrino beam, peaked at 2.5~GeV. 
The neutrino beam flux, monitored by a near detector located at Fermilab, travels through the Earth crust for 1300~km and is finally detected by a far detector installed at the Sanford Underground Research Facility in South Dakota. The far detector is constituted by at least two 17.5~kton liquid argon (LAr) Time Projection Chambers (TPC).
The huge target mass will enable a rich scientific program including among others
 searches for proton decay and  detection of the neutrino flux from a core-collapse supernova within our galaxy.
 
 LAr is known to be an excellent scintillator emitting  41 photons per keV of deposited energy by minimum ionizing particles.
 Scintillation photons are emitted through the de-excitation of Argon dimers (Ar$_{2}^{*}$) singlet (S) and triplet (T)
states with characteristic times of 6$\div$10~ns and about 1400$\div$1600~ns, respectively \cite{LAr}.
  The scintillation photons are emitted in Vacuum Ultra Violet (VUV) with a wavelength centered in a narrow band of $\sim$10~nm around 128~nm.
The primary goal of the photon detection system of the DUNE far detector must meet several requirements: convert VUV light to visible 
through the use of wavelength shifting compounds in order to make it detectable by standard (cryogenic) photo-sensitive devices; provide large coverage at reasonable cost
due to the huge detector dimensions; reach the required detection efficiency
(>1\%) to meet the supernovae DUNE scientific program. \\
The device proposed for the light detection system of the DUNE far detector is the X-ARAPUCA~\cite{Arapuca}. 
It is constituted by a light collector coupled to an array of silicon photo-multipliers (SiPMs) which detect the collected photons.
In this work the detection principle of a sample of two X-ARAPUCA devices has been probed in a continuous data-taking lasting about 10 days
with the use of a LAr condenser and of a purification system.

\section{The X-ARAPUCA}
\label{sec:XA}

\begin{figure}[h!]
\begin{center}
\includegraphics*[width=4.6cm,angle=0]{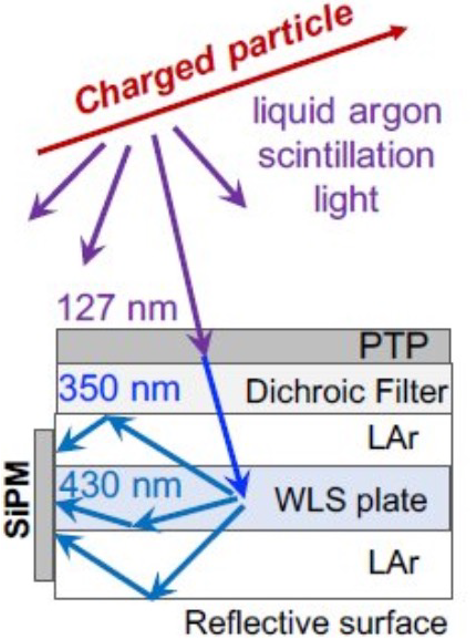}
\hspace{1cm}
\includegraphics*[width=6.2cm,angle=0]{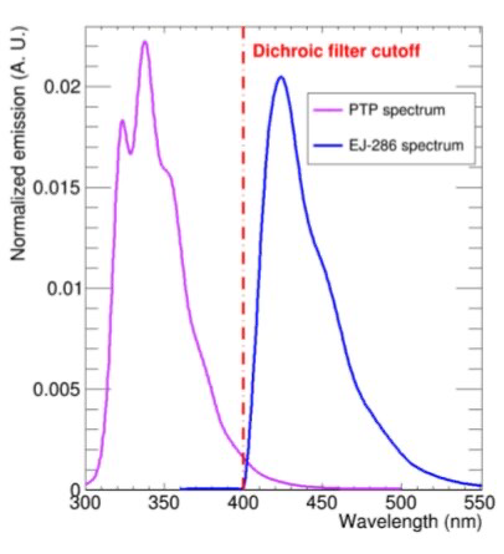}
 \caption{Left: X-ARAPUCA working principle; Right: pTP and EJ-286 wavelength shifters emission spectra.}
\label{fig:XAConcept}
\end{center}
\end{figure}

The X-ARAPUCA (XA) light trap is an evolution of the first ARAPUCA design~\cite{Arapuca2}. It consists of a 
box cavity with highly reflective internal walls. The entrance window of the box is made by a short-pass
dichroic filter which has the properties of being highly transparent to photons with wavelength below a given cut-off (400~nm),
while being highly reflective to photons with wavelength above the same cut-off.
The dichroic filter is coated on the external side of the entrance window with para-Terphenyl (pTP), a wavelength shifter 
converting photons from 128~nm to 350~nm \cite{pTP}. Because its wavelength is below the dichroic cut-off, such photons
can cross the dichroic filter. Inside the box immersed in LAr a second wavelength shifting step is performed by a WLS slab (EJ-286PS model manufactured by Eljen Technology) that shifts 350~nm photons to 430~nm. The light re-emitted from the WLS slab can be trapped by total internal reflections or escape, to be then reflected
back by the dichroic filter. In both cases the photons eventually reach the SiPM photosensor located on the edge 
of the WLS slab. The light trap sequence is summarized in fig.~\ref{fig:XAConcept}. 
The pTP and EJ-286PS emission spectra are shown in fig.~\ref{fig:XAConcept}, where the separation with respect to the dichroic cut-off is clearly visible.

%
%
%
%
%

%
%
%
%
%

\begin{figure}[h!]
\begin{center}
\includegraphics*[width=7.2cm,angle=-90]{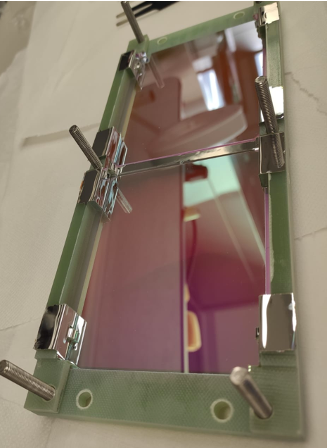}
 \caption{X-ARAPUCA device used in the test.}
\label{fig:XASBND}
\end{center}
\end{figure}

The XA concept has been realised in several shapes and sizes. The XA model used in this work features two dichroic windows sizing 200$\times$75~mm$^2$ overall,
the same format employed in SBND experiment~\cite{SBND} (fig.~\ref{fig:XASBND}). The WLS slab is coupled to four photosensor boards, each containing four SiPMs (Hamamatsu model S13360-6050VE) ganged in parallel.
A six windows XA is the basic photon detection unit for the DUNE far detector first module~\cite{DUNE-HD}.

%
%
%

\section{Experimental setup and cryogenic facility}
\label{sec:}

\begin{figure}[h!]
\begin{center}
\includegraphics*[width=5.2cm,angle=0]{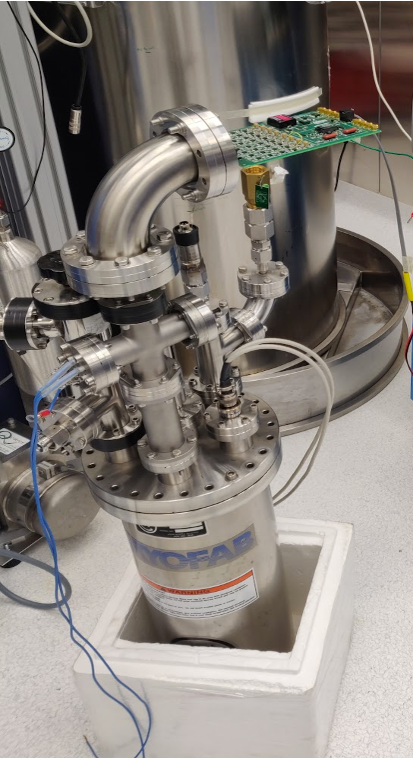}
\hspace{1cm}
\includegraphics*[width=5.8cm,angle=0]{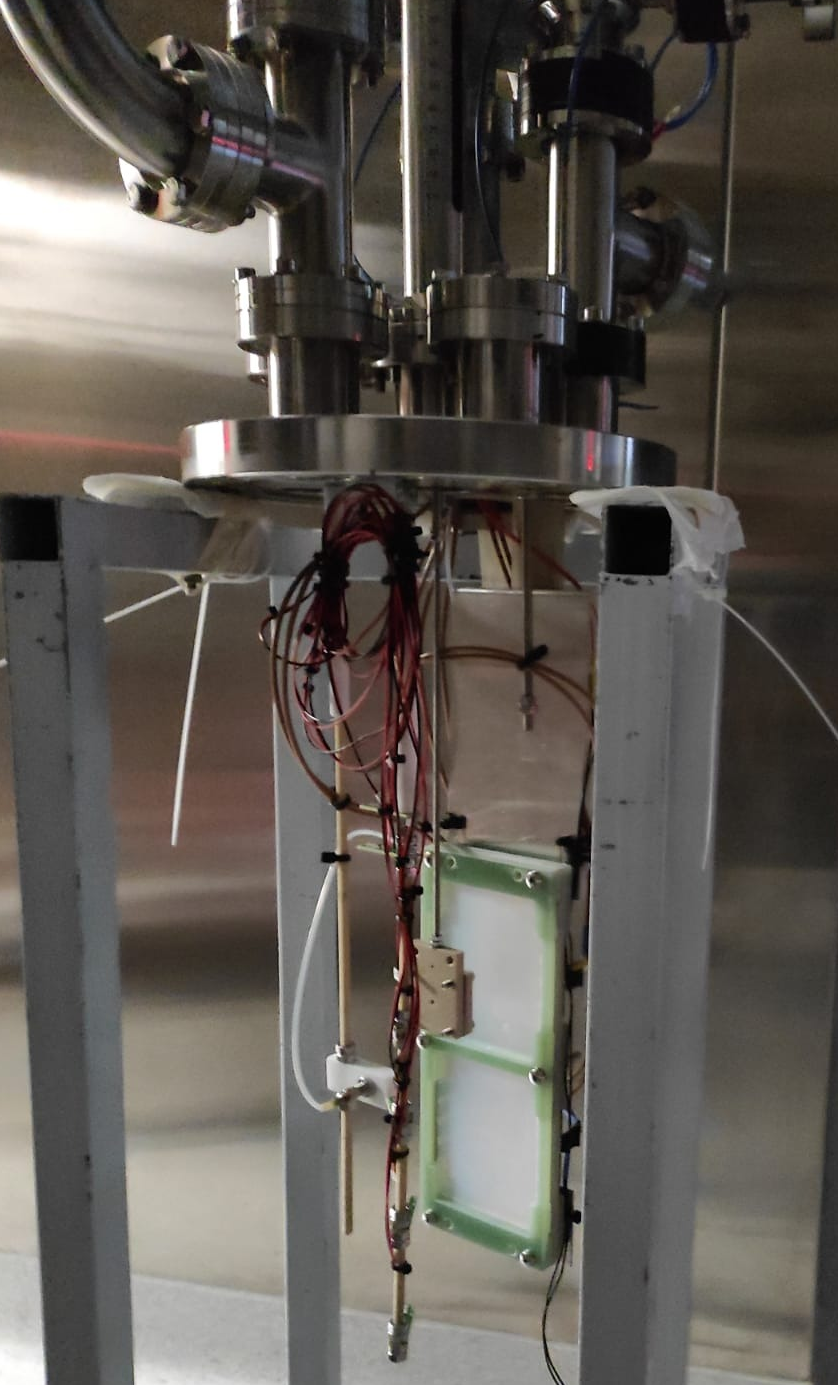}
 \caption{Cryostat hosting the X-ARAPUCA device}
\label{fig:XATest}
\end{center}
\end{figure}

The two windows XA under test was installed in a 13~l cryostat (fig.~\ref{fig:XATest}) connected to an argon gas liquefaction and purification system. The experimental setup allows for testing the photosensor performance in detecting events generated in pure liquid argon by a radioactive source.  The device is hanging from the top flange and an $^{241}$Am alpha source is mounted frontally in a peek holder fixed to a motion feedthrough. 
In this way the source can be translated between the two centers of the XA windows. The distance between the source and the dichroic surface is 4~cm.
 An optical fiber for single photoelectron calibration is inserted in the cryostat through an optical feedthrough externally connected to an Hamamatsu laser head PLP C8898.\\
The cryostat is filled by liquefying gaseous Ar~6.0 (1~ppm impurities in total) from pressurised bottles.
The argon liquefaction process is performed through a condenser made by two "brazed plate" and one "tube and shell" heat exchangers.
 The heat exchange is performed at the expenses of liquid nitrogen. 
After LAr filling the evaporated gaseous argon is recirculated by a gas pump through a  SAES MonoTorr  model PS4-MT50-R rare gas hot purifier~\cite{Getter}.
 The cryostat is equipped with six PT100 level meter sensors and a pressure transducer.\\
For this test, the four SiPMs boards from the XA were biased and read-out through the APSAIA board~\cite{Apsaia} developed within the SBND experiment. The board embodies 8 channels with input connectors. Both power supplies and amplifiers have remote control via RS232 serial port.
The four XA output channels read-out by APSAIA are then sent to CAEN V1725B digitizers (250~MS/s, 14~bit).

%
%
%
%
%

%
%
%
%
%
%
%
%
%
%
%
%
%

\section{Measurement results in Liquid Argon}
\label{sec:XATest}

\begin{figure}[h!]
\begin{center}
\includegraphics*[width=8.4cm,angle=0]{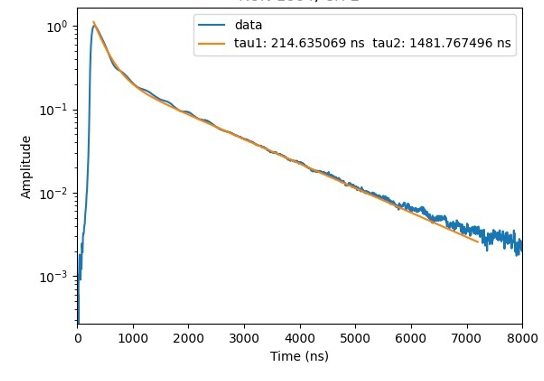}
 \caption{Average waveform from alpha source scintillation}
\label{fig:WFAlpha}
\end{center}
\end{figure}

\begin{figure}[h!]
\begin{center}
\includegraphics*[width=14.cm,angle=0]{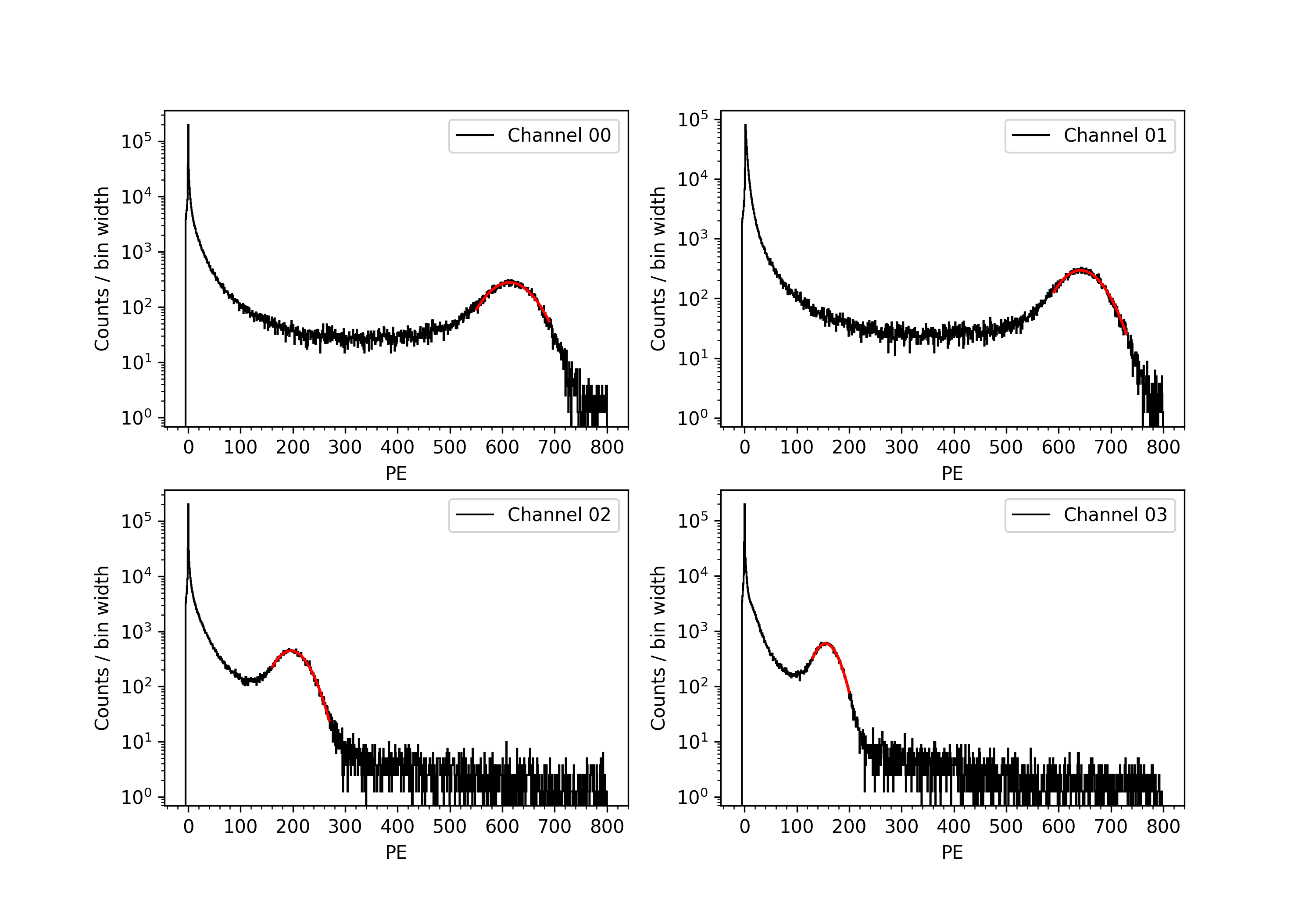}
 \caption{Alpha source spectra in PE for all four X-ARAPUCA channels}
\label{fig:AlphaSpectra}
\end{center}
\end{figure}

With SiPM bias set at 4V over-voltage, we studied the single photon response of the different channels in laser-triggered runs to calibrate the average charge of the first photo-electron peak. 
This calibration was then used to study the XA response to events generated in argon by the $^{241}$Am alpha source: a rate of about 100 Hz was found in self-trigger mode data taking.
Data presented in this work refer to the case of alpha source positioned at the center of one of the two XA windows (the one located deeper in LAr). Fig.~\ref{fig:WFAlpha} shows an example of normalized and fitted waveform from scintillation light due to alpha particles. The slow decay component is in good agreement with expected value of 1.4 ~$\mu$s.  Fig.~\ref{fig:AlphaSpectra} shows the reconstructed charge, in number of photo-electrons (PE), for all four channels. For each channel, the $\alpha$ peak is fitted with a gaussian distribution.

\subsection{Light yield and stability of the system}
\label{subsec:long_term_run} 

\begin{figure}[h!]
\begin{center}
\includegraphics*[width=11.cm,angle=0]{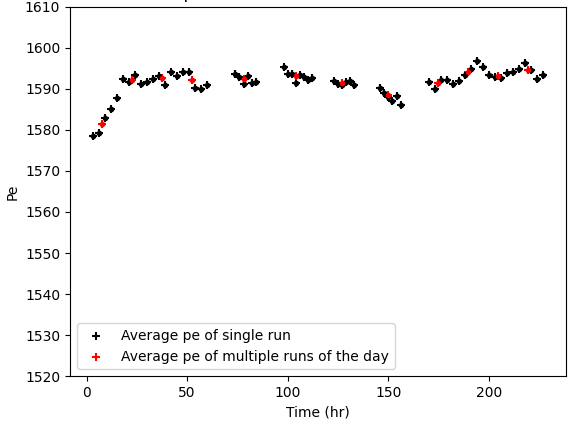}
 \caption{Total average PE from four X-ARAPUCA channels Vs time for 10 days}
\label{fig:PEStability}
\end{center}
\end{figure}

The system was kept in LAr with recirculation and purification system active for 10 consecutive days. Several self-trigger runs  were acquired daily to monitor the stability of the light yield, measured as the total average number of PEs detected by the four channels. After a slight increase in the first 20 hours of data taking, due to impurities removal, the light yield was stable during the full period within $\pm$1\%  (fig.~\ref{fig:PEStability}). The final light yield was found N$_{<PE>}$=1595$\pm$10 on average. 

\subsection{X-ARAPUCA efficiency measurement}
\label{subsec:XA_efficiency}

A preliminary efficiency of the XA photodetector can be known if the number of photons impinging on the dichroic window surface are estimated.
The number of photons produced in LAr by excitation from the alpha particles is evaluated assuming a photon yield of Y$_{\gamma}=51000\pm 1000$~photons/MeV and a quenching factor $\alpha_Q=0.71\pm 0.02$ for alpha particles~\cite{LAr, LAr2, LAr3}. With an alpha quasi-monochromatic energy of E$_{\alpha}$=5.48~MeV, the total amount of emitted photons is given by:
$$\text{N}_{\gamma}=\text{Y}_{\gamma}\times \text{E}_{\alpha}\times\alpha_{Q}$$
The geometrical acceptance for the isotropically emitted VUV photons was evaluated with a Geant4 simulation in which the XA complete geometry was implemented together with the $^{241}$Am source size and position. 
We found a geometrical efficiency $\epsilon_{geom}$=21.1\%, leading to a number of VUV photons impinging the XA surface given by:
$$\text{N}_{\gamma}^{XA}= \text{N}_{\gamma} \times \epsilon_{geom}=41300\pm 1400$$
The number of detected PEs obtained by summing all four channels must be corrected for the SiPMs secondary pulses induced by cross-talk and afterpulses. To this purpose we used the value $1.55\pm0.05$, expressed in number of avalanches generated per detected photon, reported in~\cite{ProtoDUNE} for our SiPMs. The final measured efficiency is given by
$$\epsilon_{XA}= \frac{\text{N}_{PE}^{corr}}{\text{N}_{\gamma}^{XA}}=2.5\pm0.3\%$$
where the error is dominated by systematic uncertainties on the single photo-electron calibration. The obtained value obtained is in good agreement with other performed measurements~\cite{Cattadori}.

%
%
%
%


\section{Conclusions}
\label{sec:concl}

X-ARAPUCA is the basic unit of the photon detection system of the DUNE Far Detector. In this work the test of a two-windows X-ARAPUCA device has been reported. The stability performances in liquid argon under the scintillation light produced by a 
$^{241}$Am alpha source have been investigated. A preliminary measurement of the absolute detection efficiency is also reported.

%
%
%

%



\acknowledgments
The authors would like to thank 
A. Pandalone and A Vanzanella from INFN-Napoli electronic workshop for their contribution on electronics, G. Franchi for useful support on the APSAIA read-out board.
 This work has been supported by FRA funds of Università degli Studi di Napoli "Federico II". 
 The activity during the XA test of Dr. Bergamini Machado and Prof. Segreto has been supported by FAI funds of INFN.




\newpage

\begin{thebibliography}{99}

\bibitem{Dune}
B. Abi et al., [DUNE Collaboration], \emph{Deep Underground Neutrino Experiment (DUNE) Far Detector} Technical Design Report, Volume I Introduction to DUNE, {\emph {JINST} {\bf15 08} (2020) T08008.}

\bibitem{LAr}
Doke T., \emph{Fundamental properties of liquid argon, krypton and xenon as radiation media}, {\emph {Portugal Phys}. {\bf 12} (1981) 9.}

\bibitem{LAr2}
D.-M. Mei, Z.-B. Yin, L. Stonehill and A. Hime, \emph{A model of nuclear recoil scintillation efficiency in noble liquids}, {\emph {Astroparticle Physics} {\bf 30} (2008) 12-17}.

\bibitem{LAr3} T. Doke et al., \emph{LET dependence of scintillation yields in liquid argon}, {\emph {Nuclear Instruments and
Methods in Physics Research Section A: Accelerators, Spectrometers, Detectors and Associated
Equipment} {\bf 269} (1988) 291 – 296.}


\bibitem{Arapuca}
A.A. Machado and E. Segreto, \emph{ARAPUCA a new device for liquid argon scintillation light detection},  {\emph {Journal of Instrumentation} {\bf 11} C02004 (2016)}

\bibitem{Arapuca2}
A. Machado, E. Segreto, D. Warner, A. Fauth, B. Gelli et al., \emph{The X-ARAPUCA: an improvement of the ARAPUCA device}, {\emph {Journal of Instrumentation}  {\bf 13} (2018) C04026.}

\bibitem{pTP}
T. DeVol, D. Wehe and G. Knoll, \emph{ Evaluation of p-terphenyl and 2,200 dimethyl-p-terphenyl as wavelength shifters for barium fluoride}, {\emph {Nuclear Instruments and Methods in Physics Research} Section A: Accelerators, Spectrometers, Detectors and Associated Equipment {\bf 327}
(1993) 354 Ð 362.}

\bibitem{SBND}
R. Acciarri, C. Adams, R. An, C. Andreopoulos, A. M. Ankowski, M. Antonello et al., \emph{A Proposal for a Three Detector Short-Baseline Neutrino Oscillation Program in the Fermilab Booster Neutrino Beam}, 2015.

\bibitem{DUNE-HD}
DUNE Collab., B. Abi et al.,  \emph{ Deep Underground Neutrino Experiment (DUNE), Far Detector  Technical Design Report, Volume IV: Far Detector Single-phase Technology} {\emph {JINST}  {\bf15 08} (2020) T08010.}

\bibitem{Getter}
SAES Pure Gas, Inc., PS4-MT50-R Rare Gas Purifier Product Manual.

\bibitem{Apsaia}
Apsaia: Arapuca Power Supply and Input Amplifier, Product Manual


\bibitem{ProtoDUNE}
B. Abi et al., [DUNE Collaboration], \emph{First Results on ProtoDUNE-SP liquid argon time projection chamber performance from a beam test at CERN Neutrino Platform}, {\emph {JINST} {\bf15} (2020) P120084.}

\bibitem{Cattadori}
C. Brizzolari, S. Brovelli, F. Bruni, P. Carniti, C.M. Cattadori, A. Falcone, C. Gotti, A.A. Machado, F. Meinardi, G. Pessina, E. Segreto, H.V. Souza, M. Spanu, F. Terranova and M. Torti \emph{Enhancement of the X-Arapuca photon detection device for the DUNE experiment}, {\emph {JINST} {\bf11} (2020) P090027.}




\end{thebibliography}
\end{document}